\pdfoutput=1

\documentclass[11pt]{article}

\usepackage{acl}

\usepackage{times}
\usepackage{latexsym}

\usepackage[T1]{fontenc}

\usepackage[utf8]{inputenc}

\usepackage{microtype}

\usepackage{inconsolata}
\usepackage{enumitem}
\usepackage{graphicx}
\usepackage{amsmath}
\usepackage{booktabs}
\usepackage{multirow}
\usepackage{makecell}
\usepackage{bbding}
\usepackage{pifont}
\usepackage{tablefootnote}
\usepackage{tabularx}
\usepackage{subfigure}

\newcommand{\tabincell}[2]{\begin{tabular}{@{}#1@{}}#2\end{tabular}}

\title{Prompt-Singer: Controllable Singing-Voice-Synthesis with Natural Language Prompt}

\author{Yongqi Wang\thanks{\ \ Equal contribution.}, Ruofan Hu\footnotemark[1], Rongjie Huang, Zhiqing Hong, Ruiqi Li, \\ {\bf Wenrui Liu, Fuming You, Tao Jin, Zhou Zhao} \\
Zhejiang University \\
\texttt{\{cyanbox, 3200102312\}@zju.edu.cn} 
}

\begin{document}
\maketitle
\begin{abstract}
Recent singing-voice-synthesis (SVS) methods have achieved remarkable audio quality and naturalness, yet they lack the capability to control the style attributes of the synthesized singing explicitly. We propose Prompt-Singer, the first SVS method that enables attribute controlling on singer gender, vocal range and volume with natural language. We adopt a model architecture based on a decoder-only transformer with a multi-scale hierarchy, and design a range-melody decoupled pitch representation that enables text-conditioned vocal range control while keeping melodic accuracy. Furthermore, we explore various experiment settings, including different types of text representations, text encoder fine-tuning, and introducing speech data to alleviate data scarcity, aiming to facilitate further research. Experiments show that our model achieves favorable controlling ability and audio quality. Audio samples are available at \url{http://prompt-singer.github.io}.
\end{abstract}

\section{Introduction}
\label{sec:intro}

Singing-voice-synthesis (SVS) systems \citep{chen2020hifisinger, huang2021multi, liu2022diffsinger, zhang2022visinger, zhang2022wesinger, zhang2023wesinger, hong2023unisinger}, which aim to generate high-fidelity singing voices given lyrics and pitch notes, have made significant advancements in improving audio quality and naturalness in recent years, facilitating music composition and development of entertainment industries. However, it hasn't been fully studied to control the style attributes of synthesized singing, such as speaker timbre, vocal range and energy. Despite that some works use fixed speaker IDs \cite{huang2021multi, zhang2022wesinger} or reference speech/singing segments \cite{shen2023naturalspeech, huanggenerspeech, huang2023make} to provide information on singer identity or other style attributes, these mechanisms are not user-friendly and lack the ability to control specific acoustic attributes explicitly.

An ideal approach to controlling the style of generated singing voices is to use natural language instructions as style prompts, as it can not only achieve precise control over specific attributes with certain descriptions, but also simplify user interaction, which may bring convenience to non-professional users such as musicians and video creators. However, applying natural language style prompts in singing-voice-synthesis faces several challenges:

\begin{itemize}[leftmargin=*]
    \item \textbf{Decoupling Melody and Vocal Range.} In real-life situations, different speakers (e.g. an elderly man and a little girl) may sing the same song within different vocal ranges. However, pitch annotations in SVS data are each tied to a specific singer in a certain vocal range. This coupling nature makes it challenging to generate singing voices with consistent vocal range and timbre to the prompt together with an accurate melody aligned with given pitch notes.

    \item \textbf{Textual Representation.} Despite that some works have explored connecting text representations with music, speech and general audio concepts \cite{CLAP2022, CLAP2023, wu2023large}, there is no text representation tailored for singing style descriptions, and the optimal choice of prompt representation for this task remains unknown.
    
    \item \textbf{Data Scarcity.} Due to the requirement of fine-grained annotations, existing SVS datasets \cite{liu2022diffsinger,wang2022opencpop, huang2021multi, zhang2022m4singer} are small in scale, typically consisting of only a few hours or tens of hours of singing data. This not only causes limited data diversity but also poses more challenges to learning the correlation between natural language descriptions and data distribution.
 
\end{itemize}

In this paper, we propose Prompt-Singer, the first controllable SVS model with natural language prompts to control the singer gender, vocal range and volume. Considering the outstanding performance of recent spoken LLMs \cite{borsos2023audiolm, wang2023neural, huang2023make, yang2023uniaudio} in terms of generation and in-context learning capabilities, we adopt a decoder-only transformer with a multi-scale hierarchy for conditional generation of discrete codec units of the singing, together with a unit vocoder for waveform reconstruction. To address the challenges mentioned above, we 1) design a decoupled pitch representation with a vocal range factor and a speaker-independent melody sequence, enabling voice range controlling while maintaining melodic accuracy; 2) investigate various text encoders for prompt encoding, as well as fine-tuning the encoders to seek the optimal textual representation for this task; 3) introduce speech data to alleviate data scarcity, and evaluate the model performance under different levels of low-resource singing data combined with speech data. Experiments show that our method achieves favorable style controlling accuracy on the three attributes, while keeping good audio quality and melodic accuracy. Our contributions are summarized as follows:

\begin{itemize}[leftmargin=*]
    \item We propose the first controllable SVS model with natural language prompts to control the singer gender, vocal range, and volume of the generated singing voice.
    \item We design a pitch representation for SVS that decouples voice range and melody, which enables prompt-conditioned voice range manipulation while keeping melodic accuracy.
    \item We investigate different text representations and fine-tune the text encoders to seek optimal text representation for the prompt in this task.
    \item We alleviate data scarcity by introducing speech data, which boosts prompt-SVS performances in low-resource scenarios.
\end{itemize}    
\section{Related Works}

\subsection{Singing Voice Synthesis}

Singing-voice-synthesis aims to generate human-like singing voices from lyrics and pitch notes, and recent deep-learning-based models have achieved remarkable progress in synthesized voice quality. Several works \cite{chen2020hifisinger, zhang2022wesinger, zhang2023wesinger, huang2022singgan} adopt generative adversarial networks for high-fidelity SVS. Diffsinger \cite{liu2022diffsinger} adopts a shallow diffusion mechanism to enhance the quality of the generated mel-spectrogram. VISinger \cite{zhang2022visinger} proposes an end-to-end architecture based on a variational autoencoder. UniSinger \cite{hong2023unisinger} proposes a unified framework for multiple singing-voice-related tasks based on representation disentanglement and cross-modality information matching. However, it has not been fully studied to control the style of generated singing. Previous multi-singer systems \cite{huang2021multi,zhang2022wesinger} use a fixed group of IDs to indicate singer identities. NaturalSpeech 2 \cite{shen2023naturalspeech} and Make-A-Voice \cite{huang2023make} use a reference singing or speech clip to provide holistic style information. Currently, there is a lack of fine-grained controllable methods for SVS.

\subsection{Instruct-guided Voice Generation}

Inspired by the success in text, image and audio generation guided with natural language instructions \cite{brown2020language, ramesh2021zero, kreuk2022audiogen, huang2023make3,huang2023make2, huang2023audiogpt}, some recent works have explored using text prompts to govern the stylistic attributes in voice synthesis. PromptTTS \cite{guo2023prompttts} incorporates style features from a fine-tuned BERT into a TTS backbone with attention. InstructTTS \cite{yang2023instructtts} achieves a text-controlled expressive TTS system with cross-modal representation learning. PromptTTS 2 \cite{leng2023prompttts} employs a variational network to generate reference acoustic features conditioned on text features. PromptVC \cite{yao2023promptvc} and PromptSpeaker \cite{zhang2023promptspeaker} investigate text-prompted voice conversion and speaker-embedding generation separately.  However, due to the data scarcity and the demand for precise pitch controlling, research on natural-language-instructed SVS is currently lacking.

\begin{figure*}[tb]
\includegraphics[width=\textwidth]{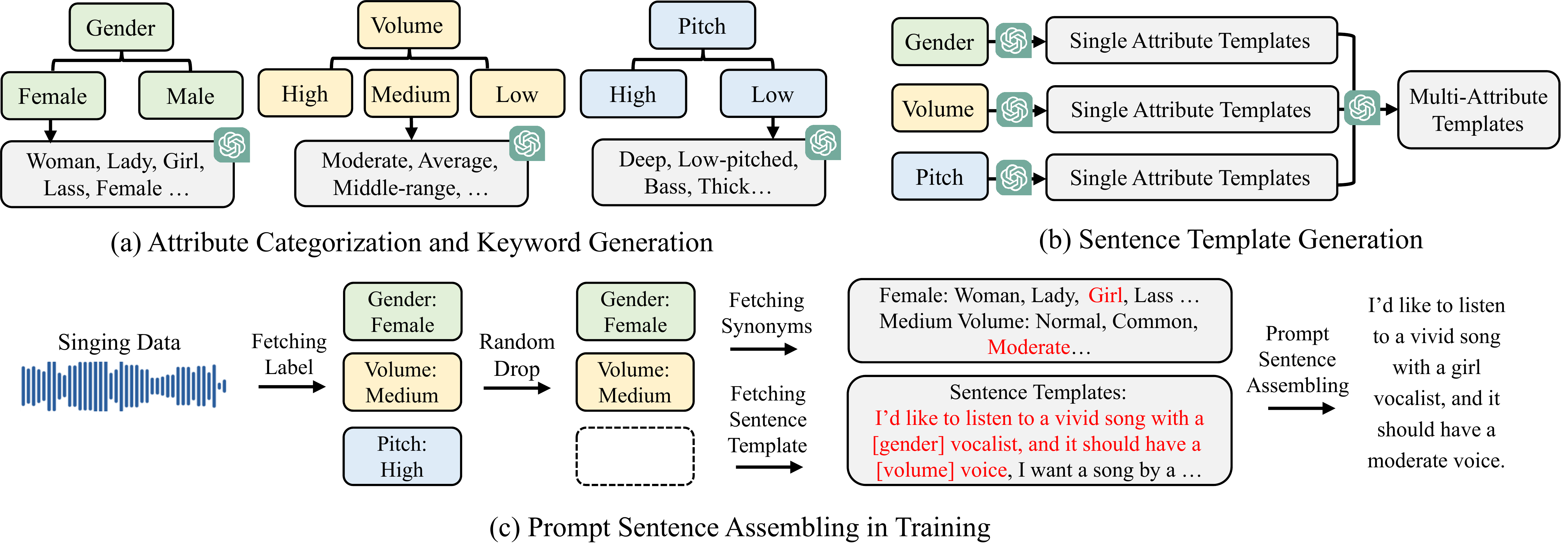}
\caption{The pipeline of generating and fetching prompt sentence for training data.}
\label{fig:prompt}
\end{figure*}

\section{Prompt Generation and Fetching}

Our goal is to control the singer gender, vocal range and volume in singing-voice-synthesis with natural language prompts. Since there is no available dataset for this task, we utilize normal SVS datasets and design a method for generating a prompt sentence for each data item. We introduce this process in this section.

Considering the high cost of manual annotation, we utilize a large language model (GPT 3.5 Turbo) to generate prompt sentences. The prompt generation mainly consists of 3 stages: 1) attribute categorization; 2) keyword and sentence template generation and 3) prompt sentence assembling.

Figure~\ref{fig:prompt}(a) and (b) demonstrate the process of the first two stages. Initially, we categorize the audio based on different attributes. The two gender categories, male and female, are pre-annotated in the datasets. For volume, we build three categories of ``low'', ``medium'', and ``high'', indicating the amplitude root mean square (RMS) ranges of $[0.02, 0.04]$, $[0.07, 0.10]$ and $[0.16, 0.20]$, respectively. Additionally, we can rescale audio into different ranges dynamically during training. For vocal range, we set two categories of ``high'' and ``low'', and use the average F0 of the voiced part as the criterion for classification, with the threshold being 125 Hz for male singers and 305 Hz for female singers.

After categorization, we use the LLM to generate a set of 4-7 synonyms for each category as the keywords. We further utilize the LLM to generate prompt sentence templates for each single attribute, where each template contains a placeholder to be replaced with the keywords (such as \textit{Generate a song by a [gender] singer}). We also generate a small number of prompt sentences targeting specific categories (such as \textit{Could you synthesize a song that's as powerful as a thunderstorm?} for large volume). We obtain approximately 50 sentence templates for each attribute after manual selection. These single-attribute templates can be further combined to create multi-attribute templates by prompting the LLM. We provide sample sentence templates and keywords in Appendix \ref{appendix:prompt}.

The prompt sentence assembling stage takes place dynamically during training. Figure~\ref{fig:prompt}(c) illustrates the pipeline of fetching a prompt sentence. We first obtain the pre-annotated labels for the data item, and in order to make the model adaptable to prompts with varying numbers of attributes, one or two labels are randomly dropped with probabilities $p_1$ and $p_2$. We then randomly fetch a keyword and a sentence template from the pre-generated sets, and replace the placeholder with the keyword to get the final prompt sentence. 
Note that we do not control vocal range independently in the absence of gender, as its boundary is different for male and female. We use pre-generated specific prompts for each sample in the evaluation for fair comparison.

\begin{figure*}[htb]
\includegraphics[width=\textwidth]{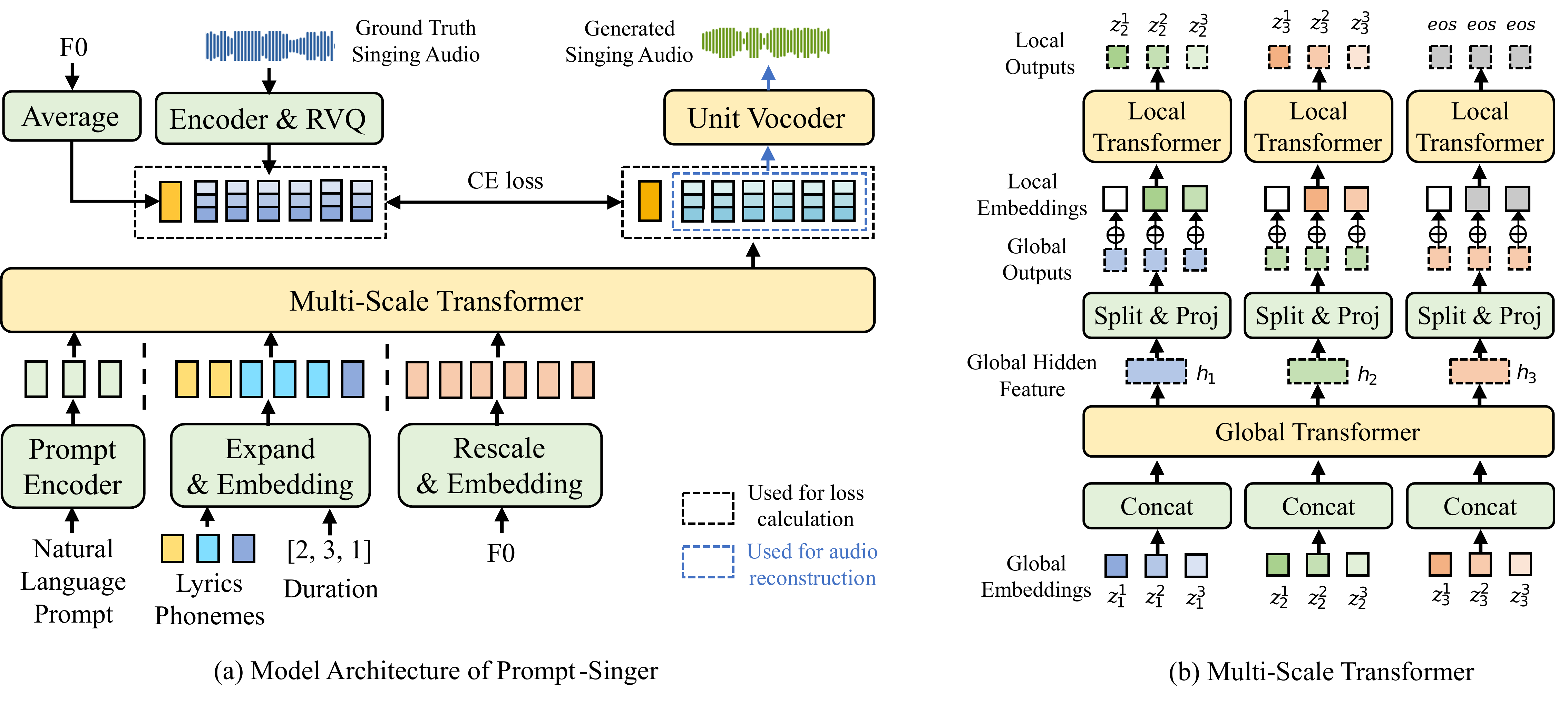}
\caption{Model architecture of Prompt-Singer and the multi-scale transformer.}
\label{fig:main}
\end{figure*}

\section{Prompt-Singer}

In this section, we introduce the model design of Prompt-Singer. The overall architecture of our model is illustrated in Figure~\ref{fig:main}(a). It is primarily composed of two sub-modules: 1) the multi-scale transformer, which generates discrete acoustic units conditioned on inputs of natural language prompt, lyrics with duration, and pitch information; and 2) the unit vocoder, which maps the generated acoustic units to an audio waveform.

In the following subsections, we introduce the input and output representations of the model in Section \ref{sec:voice_representation} to \ref{subsec:rep2}, model architecture in detail in Section \ref{subsec:multi} and \ref{subsec:vocoder}, together with our method for data scarcity alleviation in Section \ref{subsec:alle}.

\subsection{Voice Representation}
\label{sec:voice_representation}

The acoustic units used as the prediction targets of the transformer are generated by SoundStream\cite{zeghidour2021soundstream}, a neural codec with an encoder-decoder architecture and a residual vector quantizer (RVQ). Such a codec model can produce discrete compressed representations of audio by employing a convolutional encoder followed by the RVQ, and these representations can be used to reconstruct waveforms with the decoder. An acoustic unit sequence can be represented as $\mathbf{a}=[a_1^1, a_1^2, ..., a_1^C, a_2^1, ..., a_T^C], a_i^j \in \{0, 1, ..., K_a-1\},\forall 1 \leq i \leq T, 1 \leq j \leq C$, with $T, C, K_a$ being number of frames, number of residual codebooks and codebook size. 

\subsection{Textual Representation}
\label{subsec:text}
The textual input for our model comprises two components: 1) lyrics, which correspond to the content of the generated song, and 2) natural language prompt, which controls the style of the singing. We introduce their representations in this subsection. 

For lyrics, we first phonemize the text and obtain corresponding phoneme-level duration in seconds from dataset annotations or a forced-alignment tool \cite{mcauliffe2017montreal}. We then convert the duration to frame level based on a preset frame rate, and regulate the length of the phoneme sequence with this duration by duplicating phonemes. We set the frame rate of phonemes to be the same as acoustic units, making it easier for the model to learn the length alignment. The regulated phoneme sequence is then embedded by a look-up table (LUT) and fed to the transformer.

For the natural language prompt, we utilize a parameter-frozen text encoder to extract a semantic representation, followed by a linear layer for mapping its dimension to fit the transformer. To explore the impact of different text representations on style controlling, we attempt three types of encoders in our experiments: 1) BERT \cite{devlin2018bert}, a widely-used self-supervised text encoder trained with masked language modeling and next sentence prediction; 2) FLAN-T5 \cite{chung2022scaling}, the encoder of a unified text-to-text transformer fine-tuned with instructions; and 3) CLAP \cite{wu2023large}, a text encoder through contrastive pretraining on natural language and audio. 
We compare BERT and FLAN-T5 of different sizes, as well as CLAP pretrained on two different datasets. We also fine-tune BERT-large and FLAN-T5 large using prompts and corresponding labels. We fine-tune BERT with multi-label prediction and have FLAN-T5 predict the label sequence corresponding to the prompt in a text-to-text manner. Note that the prompts used in the evaluation are not included in fine-tuning.

\subsection{Decoupled Pitch Representation}
\label{subsec:rep2}
According to the equal temperament theory \cite{von1912sensations}, humans' perception of musical intervals corresponds to the logarithmic distance of frequencies. This means if we multiply the fundamental frequency (F0) of the voiced part of singing by a factor (equivalent to adding an offset in the logarithmic domain), we can adjust the vocal range without changing the melody. Based on this principle, we decompose F0 into two components: 1) $\bar{f_0}$, which is the average value of the voiced part of F0, indicting the vocal range; and 2) $\mathbf{\tilde{f_0}}=[\tilde{f_0^1}, \tilde{f_0^2}, ..., \tilde{f_0^T}]$, where we rescale the voiced part of the original F0 sequence to have a specific mean value (230Hz, in our practice), indicating vocal-range-invariant melody information. This simple yet effective representation creates an information bottleneck, forcing the model to extract melodic and vocal range information from the rescaled F0 sequence and average F0 factor, respectively. In our practice, we round $\mathbf{\tilde{f_0}}$ and $\bar{f_0}$ into integers, and use an LUT to embed them before feeding them to the transformer backbone. Both $\mathbf{\tilde{f_0}}$ and $\bar{f_0}$ share the same embedding space.

\subsection{Alleviating Data Scarcity}
\label{subsec:alle}
Considering that both speech and singing are human voices in different forms, it is intuitive that they share some commonalities in style characteristics and distributions. Based on this point, we incorporate text-to-speech (TTS) data into the training of the prompt SVS task to alleviate data scarcity. Specifically, we employ the same methods as for singing to phonemize the text and generate prompts, and use an off-the-shelf tool to extract pitch from the speech, finally obtaining data items in the same format as SVS data.

Furthermore, we explore the feasibility of substituting speech data for singing data in low-resource scenarios. We evaluate the model performance under compositions of varying amounts of low-resourced SVS data with abundant TTS data, with experiment results presented in Section \ref{subsec:res}.

\subsection{Multi-Scale Transformer Architecture}
\label{subsec:multi}

The end-to-end differentiable multi-scale transformer architecture \cite{yu2024megabyte, yang2023uniaudio} has exhibited remarkable capabilities in audio synthesis and modeling intrinsic relationships between acoustic and other modalities, as well as high efficiency of generating long sequences based on sub-quadratic self-attention. In this work, we utilize a multi-scale transformer derived from UniAudio \cite{yang2023uniaudio} to serve as the backbone of our model. It is a decoder-only transformer with a hierarchical structure to facilitate the modeling of long sequences. This module aims to generate discrete acoustic units of singing voices conditioned on natural language prompts, lyrics phonemes, phoneme durations and vocal-range agnostic melody representation, together with the vocal-range factor as intermediate output. During training, the conditional inputs and target outputs are concatenated into a single sequence and fed to the transformer, which models the correlation using next-token-prediction with cross-entropy loss calculated on the target output part. During inference, the model predicts the range factor and acoustic units conditioned on the prefix input sequence autoregressively, which can be formulated as:

\begin{small}
    \begin{gather}
    P_{cond}\left(\mathbf{a}\right) = P_{cond}\left(\bar{f_0}\right) \cdot \prod_{t=1}^T \prod_{c=1}^C P_{AR}\left(\mathbf{a}_t^c\right) \\
    P_{cond}\left(*\right) = p\left(*\mid \mathbf{E}_\mathcal{P}(\mathcal{P}), L, \mathbf{d}, \mathbf{\tilde{f_0}} ; \theta_{A R}\right) \\ 
    P_{AR}\left(\mathbf{a}_t^c\right) = p\left(\mathbf{a}_t^c \mid \mathbf{a}_{<t}, \mathbf{a}_{t}^{<c}, \mathbf{E}_\mathcal{P}(\mathcal{P}), L, \mathbf{d}, \mathbf{\tilde{f_0}}, \bar{f_0} ; \theta_{A R}\right) 
    \end{gather}
\end{small}
where $\mathbf{a}$, $\mathbf{E}_\mathcal{P}$, $\mathcal{P}$, $L$, $\mathbf{d}$, $\mathbf{\tilde{f_0}}$, $\bar{f0}$ and $\theta_{A R}$ indicate acoustic units, prompt encoder, prompt, lyrics, durations, melody representation, vocal-range factor and model parameters, respectively, and $t$, $c$ indicate temporal and codebook indices of the acoustic unit. Consider the process of the transformer predicting the vocal range factor, which is formulated by

\begin{small}
\begin{equation}
    P_{cond}\left(\bar{f_0}\right)= p\left(\bar{f_0} \mid \mathbf{E}_\mathcal{P}(\mathcal{P}), L, \mathbf{d}, \mathbf{\tilde{f_0}} ; \theta_{A R}\right) ,
\end{equation}
\end{small}
as we assume that the average F0 value is independent of the lyrics, duration and melody, this formula indicates our model's capability to control the vocal range through natural language prompts. The predicted vocal range information is further taken as a condition for singing acoustic unit generation.

The hierarchical structure of the multi-scale transformer is illustrated in Figure~\ref{fig:main}(b). This structure is formed by a global and a local transformer, both of which are decoder-only transformers. For a temporal position $t$, embeddings $z^{1:n_q}_t$ of acoustic units from different codebooks are concatenated and fed to the global transformer for inter-frame correlation modeling. The output hidden feature $h_t$ is generated autoregressively conditioned on $h_{1:t-1}$. This hidden feature is then split according to the original shape of the embeddings, projected by a linear layer, and added to the input embeddings of the local transformer as a frame-level context. The local transformer predicts acoustic units of different codebooks inside a frame autoregressively. For non-acoustic modalities, each item is repeated $n_q$ times to fit this modeling mechanism, with $n_q$ being the number of codebooks.

\subsection{Unit Vocoder}
\label{subsec:vocoder}
When the acoustic unit generation finishes, the generated units need to be mapped to a high-fidelity audio waveform. Due to the compressive nature of the codec, reconstructing audio from acoustic units of limited codebooks with the decoder may result in degraded perceptual quality. Instead of using the codec decoder directly, we adopt a GAN-based unit vocoder for singing voice reconstruction, aiming to generate audio of higher quality and richer details. Specifically, our vocoder is derived from BigVGAN \cite{lee2022bigvgan}, with a generator built from a set of LUTs that embed the discrete units, and a series of blocks composed of transposed convolution and a residual block with dilated layers. Multi-period and multi-resolution discriminators (MPD, MRD) are used for adversarial training. 

\section{Experiments}
\label{sec:exp}

\subsection{Datasets}

We combine 4 SVS datasets for our task, including M4Singer, Opencpop, Opensinger and PopCS, forming a multi-singer singing dataset of 127 hours. For speech data, we utilize 4 Mandarin TTS corpora, including AISHELL-3, Biaobei, THCHS-30 and a subset of DidiSpeech, totaling approximately 179 hours. We provide details of these datasets in Appendix \ref{appendix:dataset}.

We phonemize the lyrics with PyPinyin\footnote{https://github.com/mozillazg/python-pinyin}, and extract F0 from raw audios with harvest \cite{morise2017harvest}. 
 
\subsection{Model Configurations}

The global transformer has 20 layers with 320M parameters, while the local transformer has 6 layers with 100M parameters. Both of them share the same hidden dimension of 1152. For acoustic units, we train a SoundStream model for 24k audio, with 12 quantization levels, a codebook size of 1024 and a downsampling rate of 480. We use the first 3 quantization levels as the acoustic units, and the unit vocoder is trained to reconstruct 24k audios from acoustic units of 3 codebooks. The label dropping probability $p_1$ and $p_2$ are both set to 0.05. Detailed structure and hyper-parameters of the model are appended in Appendix \ref{appendix:model}.

\subsection{Experiment Settings}
\label{subsec:setting}

\begin{table*}[ht]
    \small
    \centering
    \begin{tabular}{clcccccc}
    \toprule
    ID & Model  & Gender (F/M) & Volume & Range & R-FFE & MOS & RMOS   \\ 
    \midrule
    \multicolumn{6}{l}{\textbf{Prompt-Singer with Pre-trained Text Encoders}} \\
    \midrule
    1 & FLAN-T5 small & 76.7 / 78.1 & 92.0 & 79.1 & 0.11 & 3.75 $\pm$ 0.08 & 3.27 $\pm$ 0.09 \\
    2 & FLAN-T5 base & 82.2 / 79.5 & 92.4 & 80.8 & 0.12 & 3.79 $\pm$ 0.07 & 3.39 $\pm$ 0.07 \\
    3 & FLAN-T5 large & 83.1 / 80.8 & 92.7 & 82.6 & 0.12 & 3.83 $\pm$ 0.08 & 3.43 $\pm$ 0.08 \\
    4 & FLAN-T5 XL & 83.4 / 80.4 & 92.6 & 82.9 & 0.11 & 3.84 $\pm$ 0.06 & 3.46 $\pm$ 0.08 \\
    5 & BERT-base & 80.8 / 80.1 & 93.9 & 80.1 & 0.10 & 3.81 $\pm$ 0.06 & 3.42 $\pm$ 0.07 \\
    6 & BERT-large & 84.9 / 80.9 & 94.3 & 78.9 & \textbf{0.09} & 3.78 $\pm$ 0.08 & 3.44 $\pm$ 0.08 \\
    7 & CLAP-general & 82.2 / 79.5 & 94.1 & 80.3 & 0.12 & 3.83 $\pm$ 0.07 & 3.43 $\pm$ 0.06 \\ 
    8 & CLAP-speech/music & 82.2 / 78.1 & 94.2 & 80.8 & 0.11 & 3.85 $\pm$ 0.09 & 3.38 $\pm$ 0.08 \\
    \midrule
    \multicolumn{6}{l}{\textbf{Prompt-Singer with Fine-tuned Text Encoders}} \\
    \midrule
    9 & FLAN-T5 large finetuned & \textbf{87.7} / \textbf{86.3} & 94.4 & \textbf{84.7} & 0.12 & 3.89 $\pm$ 0.07 & \textbf{3.62 $\pm$ 0.08}  \\
    10 & BERT-large finetuned & 86.3 / 83.6 & \textbf{94.9} & 79.8 & 0.10 & \textbf{3.90 $\pm$ 0.07} & 3.60 $\pm$ 0.08 \\ 
    \midrule
    \multicolumn{6}{l}{\textbf{Non-controllable SVS models and Ground Truth}} \\
    \midrule
    11 & FFT-Singer & / & / & / & 0.17 &  3.67 $\pm$ 0.08 & / \\
    12 & Diffsinger & / & / & / &  \textbf{0.09} &  3.86 $\pm$ 0.07  &  / \\
    13 & Ground Truth & 98.0 / 97.0 & / & / & / &  4.09 $\pm$ 0.06 & / \\
    \bottomrule 
    \end{tabular}
    \caption{Results on different text representations, including percentage accuracies of the three attributes, rescaled f0-frame error (R-FFE) and mean-opinion-scores of audio quality (MOS) and relevance to the prompt (RMOS).}
    \label{tab:res}
\end{table*}

\begin{table*}[ht]
    \small
    \centering
    \begin{tabular}{ccccccccc}
    \toprule
    ID & SVS Data & TTS Data & Gender (F/M) & Volume & Range & R-FFE & MOS & RMOS  \\ 
    \midrule
    1 & \ding{51} & \ding{55} & 75.3 / 65.8 & 87.6 & 78.7 & 0.11 & 3.68 $\pm$ 0.08 & 3.37 $\pm$ 0.08 \\
    2 & \ding{51} & \ding{51} & 87.7 / 86.3 & 94.4 & 84.7 & 0.12 & 3.89 $\pm$ 0.07 & 3.62 $\pm$ 0.08 \\
    \midrule
    3 & 10min & 100h & 65.8 / 65.6 & 78.3 & 80.9 & 0.29 & 3.06 $\pm$ 0.09 & 2.89 $\pm$ 0.09  \\
    4 & 1h & 100h & 71.2 / 64.4 & 84.8 & 81.2 & 0.25 & 3.34 $\pm$ 0.08 & 3.03 $\pm$ 0.09  \\ 
    5 & 10h & 100h &  76.7 / 68.5 & 88.6 & 81.6 & 0.23 & 3.28 $\pm$ 0.08 & 3.17 $\pm$ 0.09 \\ 
    6 & 100h & 100h  & 86.2 / 80.5 & 92.5 & 82.3 & 0.12 & 3.75 $\pm$ 0.08 & 3.45 $\pm$ 0.08 \\
    \bottomrule 
    \end{tabular}
    \caption{Experiment results on data scarcity alleviation in low resource scenarios.}
    \label{tab:res2}
\end{table*}

As we are investigating a new task with no previous work to compare with, our experiments mainly focus on exploring different settings within our framework, including different text representations and different training data compositions, together with ablation studies. The settings of various text representations are presented in table \ref{tab:res}. As described in Section \ref{subsec:text}, we experimented with encoders of different types, parameter sizes, and pre-training data as well as fine-tuning the encoders. We also provide the results of ground truth and two non-controllable SVS models in table \ref{tab:res} as baselines of singing quality: 1) FFT-Singer, which generates mel-spectrograms through stacked feed-forward transformer blocks; and 2) Diffsinger\cite{liu2022diffsinger}, an SVS model based on the diffusion probabilistic model. 

In table \ref{tab:res2}, we compare the results of incorporating speech data for training or not, together with a series of low-resource data configurations with SVS data varying from 10 minutes to 100 hours paired with speech data of a fixed quantity of 100 hours. The ablation studies are described in a dedicated subsection.

\subsection{Metrics}

We employ both subjective and objective metrics to measure the controlling ability and singing voice quality of the models. For objectives metrics, we calculate the percentage accuracy for each attribute, where we train a gender classifier and use amplitude RMS and average F0 of the voiced part for volume and range evaluation. We mainly use single-attribute prompts for evaluation with an additional gender attribute for vocal range, and multi-attribute evaluation is conducted in ablation studies. We also calculate R-FFE for melodic accuracy between the synthesized and reference singing, which is F0-frame-error (FFE) with the voiced part of F0 rescaled to have the same average value for both singing segments to eliminate the impact of vocal range, where the new average value is the mean of the original means of the two segments. For subjective metrics, we use crowd-sourced human evaluation via Amazon Mechanical Turk, where raters are asked to rate scores on 1-5 Likert scales on singing voice quality and the relevance between synthesized singing and the prompt. 
We report the mean-opinion-scores of quality (MOS) and relevance (RMOS) with 95\% confidence intervals (CI) in the tables. Details of evaluation metrics are provided in Appendix \ref{appendix:eval}.

\subsection{Results and Analysis}
\label{subsec:res}

We can draw a basic conclusion from the results in table \ref{tab:res}: generally, our models (1-10) exhibit favorable attribute controlling accuracies, with the best values being 87.7 / 86.3, 94.9 and 84.7 for the three attributes, together with competitive audio quality and melodic accuracy to non-controllable baselines (1-10 v.s. 11-13), with the best R-FFE and MOS being 0.09 and 3.90. This indicates the effectiveness of our model design on the task of controllable SVS. 

\begin{table*}[ht]
    \small
    \centering
    \begin{tabular}{clccccc}
    \toprule
     ID & Model & Gender (F/M) & Volume & Range & R-FFE & RMOS   \\
     \midrule
     \multicolumn{7}{l}{Ablation on Decoupled Pitch Representation} \\
     \midrule
    1 & Factor: \ding{51}  Rescale: \ding{51} & 87.7 / 86.3 & 94.4 & 84.7 & 0.12 & 3.62 $\pm$ 0.08 \\
    2 & Factor: \ding{55} \  Rescale: \ding{51} & 78.1 / 63.0 & 91.3 & 76.1  & 0.11 & 3.34 $\pm$ 0.09 \\ 
    3 & Factor: \ding{55} \  Rescale: \ding{55} & 64.4 / 58.9 & 91.6 & 72.3 & 0.08 & 2.75 $\pm$ 0.09 \\ 
    \midrule
     \multicolumn{7}{l}{Ablation on Different Prompted Attribute Numbers} \\ 
    \midrule
    4 & Attribute Num: 1 &  87.7 / 86.3 & 94.4 & / & 0.12 & 3.67 $\pm$ 0.08\\
    5 & Attribute Num: 2 & 84.3 / 82.9 & 93.4 &  84.7 & 0.11 & 3.58 $\pm$ 0.08 \\
    6 & Attribute Num: 3 & 81.2 / 80.7 & 93.0 & 82.4 & 0.11 & 3.52 $\pm$ 0.07 \\
    \bottomrule 
    \end{tabular}
    \caption{Results of ablation studies.}
    \label{tab:abl}
\end{table*}

\subsubsection{Evaluation on Text Representations}

We have the following further observations from the results in table \ref{tab:res}: 1) Fine-tuning the text encoders leads to a considerable improvement in controlling accuracy (3 vs. 9 \& 6 vs.10), with the improvements being 4.6 / 5.5, 1.7 and 2.1 for FLAN-T5 large, and 1.4 / 2.7, 0.6 and 0.9 for BERT-large. This indicates that aligning the text representations with the labels, which have a much simpler distribution, helps the model learn their correlation with singing style. Nevertheless, using only the pre-trained text encoders already yields quite good results. 2) Generally, larger model sizes bring better results (1-4 \& 5-6). However, such a tendency between 3 and 4 is less significant compared to 1-2 and 2-3, suggesting that text encoder parameters beyond a certain size are no longer a bottleneck for model performance. 3) 
Different types of text encoders exhibit varying controlling capabilities over different attributes. For instance (1-4 vs. 5-8), the FLAN-T5 family shows weaker control over volume compared to CLAP and BERT, with an accuracy gap of 1.2-2.3. However, the large and xl models outperform CLAP and BERT in vocal-range controlling accuracy by 1.8-4.0. 
This may be related to differences in the models' pretraining methods and data. We choose the fine-tuned FLAN-T5 large model for subsequent experiments. 

\subsubsection{Evaluation on Data Scarcity Alleviation}

From the results of different data compositions in table \ref{tab:res2}, we have the following observations: 1) Introducing speech data leads to a comprehensive improvement in controlling accuracies and generation quality, with the cost being a slight increase in R-FFE of 0.01 (1 vs. 2). This is because the additional speech data increases the quantity and diversity of the training data, aiding the network in modeling the correlation between prompt and acoustic style. However, due to the difference in the distributions of singing melody and speech prosody, both of which are manifested in pitch variation, the speech data may have a negative impact on modeling singing melody, causing the slight increase in R-FFE. 2) In the low resource scenarios (3-6), we find that there is a drastic decline in the singing audio quality, melody accuracy as well as the accuracy on gender with the decrease in the quantity of SVS data. In contrast, the changes in volume and vocal range are relatively gradual, yielding acceptable results of 88.6 and 81.6 even with 10 hours of singing data. This suggests that, while speech data helps improve controlling accuracy and audio quality, it still cannot substitute for singing data in modeling certain vocal characteristics. In conclusion, introducing speech data effectively enhances the performance of controllable SVS, but it is still necessary to have a sufficient amount of singing data to ensure synthesis quality and melody accuracy.

\subsection{Ablation Studies}

We mainly focus on validating the effectiveness of our decoupled pitch representation and multi-attribute prompting mechanism in the ablation studies, and the results are presented in table \ref{tab:abl}.

For pitch representation (1-3), we first remove the vocal range factor from the sequence, and then eliminate the rescaling on the input F0. We can see that when removing the range factor, there is a drastic drop of 9.6 / 23.3, 3.1 and 8.6 in accuracies, accompanied by an RMOS decrease of 0.28. This indicates that explicitly predicting the vocal range factor facilitates vocal range and gender control greatly. When we continue to eliminate the input F0 rescaling, the accuracies on gender and range as well as RMOS further decline by 13.7 / 4.1, 3.8 and 0.59, respectively, which indicates that the vocal range information contained in the original F0 interferes with the model's modeling of the correlation between prompt and singing style. We also observe that removing the range factor and input F0 rescaling leads to an improvement in melodic accuracy. This suggests that the decoupling mechanism may cause some loss of pitch information. Despite this, our model keeps a satisfactory melodic accuracy with the decoupled pitch representation.

We further examine the model's controlling effectiveness under multi-attribute prompts. The results of 4-6 in table \ref{tab:abl} show that there is a slight decrease in accuracies and RMOS as the attribute number increases, with the drop being 3.4 / 3.4, 1.0, 0.09 from 1 to 2 attributes, and 3.1 / 2.2, 0.4, 2.3, 0.06 from 2 to 3. We suggest that this is because the conditional distribution of acoustic style with respect to controlling signals of multiple attributes is more complicated to be modeled. Nevertheless, our model shows favorable performance on prompts with both single and multiple attributes.

\section{Conclusion}
In this paper, we propose Prompt-Singer, the first singing-voice-synthesis method with the ability of style control using natural language prompts. We adopt a multi-scale decoder-only transformer for generating acoustic units of singing, followed by a unit-vocoder for audio reconstruction. We design a decoupled pitch representation for vocal range modification with an accurate melody kept. Furthermore, we investigate various experiment settings, including different text representations, fine-tuning the text encoders, and using speech data to boost performance in low-resource scenarios. 

In future works, we plan to introduce more style attributes in controllable SVS, such as emotion, rhythm and more detailed singer information. We hope our work will facilitate the development of the SVS community.

\section{Limitations and Potential Risks}
Despite that our model achieves remarkable controlling capability and audio quality on prompt singing-voice-synthesis, it still has two major limitations: 1) Due to the simplicity and inflexibility of our existing prompt generation pipeline, the generated prompt texts may suffer from distributional bias, manifested mainly as grammatical errors, unnatural expressions, and restrictions in expressive capacity and diversity. We suggest that a potential solution is to pass the assembled prompt sentences through the LLM once more for refinement and synonymous sentence generation to improve accuracy and expressiveness. 2) Due to the utilization of large-scale models (including the text encoders and the transformer backbone) along with an autoregressive generation paradigm, our model entails relatively high computational overhead, resulting in considerable inference latency. We discuss the relationship between inference latency and the length of the generated audio in appendix \ref{appendix:latency}.

Besides, misuse of our model for singing voice generation may lead to copyright issues. We will add some constraints to guarantee people who use our code or pre-trained model will not use the model in illegal cases.

\label{sec:bibtex}

\section*{Acknowledgements}
This work is supported by National Key R\&D Program of China under Grant No.2022ZD0162000, National Natural Science Foundation of China under Grant No. 62222211 and No.62072397. 

\bibliography{anthology,custom}

\begin{thebibliography}{41}
\expandafter\ifx\csname natexlab\endcsname\relax\def\natexlab#1{#1}\fi

\bibitem[{Borsos et~al.(2023)Borsos, Marinier, Vincent, Kharitonov, Pietquin, Sharifi, Roblek, Teboul, Grangier, Tagliasacchi et~al.}]{borsos2023audiolm}
Zal{\'a}n Borsos, Rapha{\"e}l Marinier, Damien Vincent, Eugene Kharitonov, Olivier Pietquin, Matt Sharifi, Dominik Roblek, Olivier Teboul, David Grangier, Marco Tagliasacchi, et~al. 2023.
\newblock Audiolm: a language modeling approach to audio generation.
\newblock \emph{IEEE/ACM Transactions on Audio, Speech, and Language Processing}.

\bibitem[{Brown et~al.(2020)Brown, Mann, Ryder, Subbiah, Kaplan, Dhariwal, Neelakantan, Shyam, Sastry, Askell et~al.}]{brown2020language}
Tom Brown, Benjamin Mann, Nick Ryder, Melanie Subbiah, Jared~D Kaplan, Prafulla Dhariwal, Arvind Neelakantan, Pranav Shyam, Girish Sastry, Amanda Askell, et~al. 2020.
\newblock Language models are few-shot learners.
\newblock \emph{Advances in neural information processing systems}, 33:1877--1901.

\bibitem[{Chen et~al.(2020)Chen, Tan, Luan, Qin, and Liu}]{chen2020hifisinger}
Jiawei Chen, Xu~Tan, Jian Luan, Tao Qin, and Tie-Yan Liu. 2020.
\newblock Hifisinger: Towards high-fidelity neural singing voice synthesis.
\newblock \emph{arXiv preprint arXiv:2009.01776}.

\bibitem[{Chung et~al.(2022)Chung, Hou, Longpre, Zoph, Tay, Fedus, Li, Wang, Dehghani, Brahma et~al.}]{chung2022scaling}
Hyung~Won Chung, Le~Hou, Shayne Longpre, Barret Zoph, Yi~Tay, William Fedus, Yunxuan Li, Xuezhi Wang, Mostafa Dehghani, Siddhartha Brahma, et~al. 2022.
\newblock Scaling instruction-finetuned language models.
\newblock \emph{arXiv preprint arXiv:2210.11416}.

\bibitem[{Devlin et~al.(2018)Devlin, Chang, Lee, and Toutanova}]{devlin2018bert}
Jacob Devlin, Ming-Wei Chang, Kenton Lee, and Kristina Toutanova. 2018.
\newblock Bert: Pre-training of deep bidirectional transformers for language understanding.
\newblock \emph{arXiv preprint arXiv:1810.04805}.

\bibitem[{Dong~Wang(2015)}]{THCHS30_2015}
Zhiyong~Zhang Dong~Wang, Xuewei~Zhang. 2015.
\newblock \href {http://arxiv.org/abs/1512.01882} {Thchs-30 : A free chinese speech corpus}.

\bibitem[{Elizalde et~al.(2023{\natexlab{a}})Elizalde, Deshmukh, Al~Ismail, and Wang}]{CLAP2022}
Benjamin Elizalde, Soham Deshmukh, Mahmoud Al~Ismail, and Huaming Wang. 2023{\natexlab{a}}.
\newblock Clap learning audio concepts from natural language supervision.
\newblock In \emph{ICASSP 2023-2023 IEEE International Conference on Acoustics, Speech and Signal Processing (ICASSP)}, pages 1--5. IEEE.

\bibitem[{Elizalde et~al.(2023{\natexlab{b}})Elizalde, Deshmukh, and Wang}]{CLAP2023}
Benjamin Elizalde, Soham Deshmukh, and Huaming Wang. 2023{\natexlab{b}}.
\newblock \href {http://arxiv.org/abs/2309.05767} {Natural language supervision for general-purpose audio representations}.

\bibitem[{Guo et~al.(2021)Guo, Wen, Jiang, Luo, Zhang, Zhao, Li, Gong, Zou, Han et~al.}]{guo2021didispeech}
Tingwei Guo, Cheng Wen, Dongwei Jiang, Ne~Luo, Ruixiong Zhang, Shuaijiang Zhao, Wubo Li, Cheng Gong, Wei Zou, Kun Han, et~al. 2021.
\newblock Didispeech: A large scale mandarin speech corpus.
\newblock In \emph{ICASSP 2021-2021 IEEE International Conference on Acoustics, Speech and Signal Processing (ICASSP)}, pages 6968--6972. IEEE.

\bibitem[{Guo et~al.(2023)Guo, Leng, Wu, Zhao, and Tan}]{guo2023prompttts}
Zhifang Guo, Yichong Leng, Yihan Wu, Sheng Zhao, and Xu~Tan. 2023.
\newblock Prompttts: Controllable text-to-speech with text descriptions.
\newblock In \emph{ICASSP 2023-2023 IEEE International Conference on Acoustics, Speech and Signal Processing (ICASSP)}, pages 1--5. IEEE.

\bibitem[{Hong et~al.(2023)Hong, Cui, Huang, Zhang, Liu, He, and Zhao}]{hong2023unisinger}
Zhiqing Hong, Chenye Cui, Rongjie Huang, Lichao Zhang, Jinglin Liu, Jinzheng He, and Zhou Zhao. 2023.
\newblock Unisinger: Unified end-to-end singing voice synthesis with cross-modality information matching.
\newblock In \emph{Proceedings of the 31st ACM International Conference on Multimedia}, pages 7569--7579.

\bibitem[{Huang et~al.(2023{\natexlab{a}})Huang, Ren, Huang, Yang, Ye, Zhang, Liu, Yin, Ma, and Zhao}]{huang2023make3}
Jiawei Huang, Yi~Ren, Rongjie Huang, Dongchao Yang, Zhenhui Ye, Chen Zhang, Jinglin Liu, Xiang Yin, Zejun Ma, and Zhou Zhao. 2023{\natexlab{a}}.
\newblock Make-an-audio 2: Temporal-enhanced text-to-audio generation.
\newblock \emph{arXiv preprint arXiv:2305.18474}.

\bibitem[{Huang et~al.(2021)Huang, Chen, Ren, Liu, Cui, and Zhao}]{huang2021multi}
Rongjie Huang, Feiyang Chen, Yi~Ren, Jinglin Liu, Chenye Cui, and Zhou Zhao. 2021.
\newblock Multi-singer: Fast multi-singer singing voice vocoder with a large-scale corpus.
\newblock In \emph{Proceedings of the 29th ACM International Conference on Multimedia}, pages 3945--3954.

\bibitem[{Huang et~al.(2022)Huang, Cui, Chen, Ren, Liu, Zhao, Huai, and Wang}]{huang2022singgan}
Rongjie Huang, Chenye Cui, Feiyang Chen, Yi~Ren, Jinglin Liu, Zhou Zhao, Baoxing Huai, and Zhefeng Wang. 2022.
\newblock Singgan: Generative adversarial network for high-fidelity singing voice generation.
\newblock In \emph{Proceedings of the 30th ACM International Conference on Multimedia}, pages 2525--2535.

\bibitem[{Huang et~al.(2023{\natexlab{b}})Huang, Huang, Yang, Ren, Liu, Li, Ye, Liu, Yin, and Zhao}]{huang2023make2}
Rongjie Huang, Jiawei Huang, Dongchao Yang, Yi~Ren, Luping Liu, Mingze Li, Zhenhui Ye, Jinglin Liu, Xiang Yin, and Zhou Zhao. 2023{\natexlab{b}}.
\newblock Make-an-audio: Text-to-audio generation with prompt-enhanced diffusion models.
\newblock In \emph{International Conference on Machine Learning}, pages 13916--13932. PMLR.

\bibitem[{Huang et~al.(2023{\natexlab{c}})Huang, Li, Yang, Shi, Chang, Ye, Wu, Hong, Huang, Liu et~al.}]{huang2023audiogpt}
Rongjie Huang, Mingze Li, Dongchao Yang, Jiatong Shi, Xuankai Chang, Zhenhui Ye, Yuning Wu, Zhiqing Hong, Jiawei Huang, Jinglin Liu, et~al. 2023{\natexlab{c}}.
\newblock Audiogpt: Understanding and generating speech, music, sound, and talking head.
\newblock \emph{arXiv preprint arXiv:2304.12995}.

\bibitem[{Huang et~al.()Huang, Ren, Liu, Cui, and Zhao}]{huanggenerspeech}
Rongjie Huang, Yi~Ren, Jinglin Liu, Chenye Cui, and Zhou Zhao.
\newblock Generspeech: Towards style transfer for generalizable out-of-domain text-to-speech.
\newblock In \emph{Advances in Neural Information Processing Systems}.

\bibitem[{Huang et~al.(2023{\natexlab{d}})Huang, Zhang, Wang, Yang, Liu, Ye, Jiang, Weng, Zhao, and Yu}]{huang2023make}
Rongjie Huang, Chunlei Zhang, Yongqi Wang, Dongchao Yang, Luping Liu, Zhenhui Ye, Ziyue Jiang, Chao Weng, Zhou Zhao, and Dong Yu. 2023{\natexlab{d}}.
\newblock Make-a-voice: Unified voice synthesis with discrete representation.
\newblock \emph{arXiv preprint arXiv:2305.19269}.

\bibitem[{Kreuk et~al.(2022)Kreuk, Synnaeve, Polyak, Singer, D{\'e}fossez, Copet, Parikh, Taigman, and Adi}]{kreuk2022audiogen}
Felix Kreuk, Gabriel Synnaeve, Adam Polyak, Uriel Singer, Alexandre D{\'e}fossez, Jade Copet, Devi Parikh, Yaniv Taigman, and Yossi Adi. 2022.
\newblock Audiogen: Textually guided audio generation.
\newblock \emph{arXiv preprint arXiv:2209.15352}.

\bibitem[{Lee et~al.(2022)Lee, Ping, Ginsburg, Catanzaro, and Yoon}]{lee2022bigvgan}
Sang-gil Lee, Wei Ping, Boris Ginsburg, Bryan Catanzaro, and Sungroh Yoon. 2022.
\newblock Bigvgan: A universal neural vocoder with large-scale training.
\newblock \emph{arXiv preprint arXiv:2206.04658}.

\bibitem[{Leng et~al.(2023)Leng, Guo, Shen, Tan, Ju, Liu, Liu, Yang, Zhang, Song et~al.}]{leng2023prompttts}
Yichong Leng, Zhifang Guo, Kai Shen, Xu~Tan, Zeqian Ju, Yanqing Liu, Yufei Liu, Dongchao Yang, Leying Zhang, Kaitao Song, et~al. 2023.
\newblock Prompttts 2: Describing and generating voices with text prompt.
\newblock \emph{arXiv preprint arXiv:2309.02285}.

\bibitem[{Liu et~al.(2022)Liu, Li, Ren, Chen, and Zhao}]{liu2022diffsinger}
Jinglin Liu, Chengxi Li, Yi~Ren, Feiyang Chen, and Zhou Zhao. 2022.
\newblock Diffsinger: Singing voice synthesis via shallow diffusion mechanism.
\newblock In \emph{Proceedings of the AAAI conference on artificial intelligence}, volume~36, pages 11020--11028.

\bibitem[{McAuliffe et~al.(2017)McAuliffe, Socolof, Mihuc, Wagner, and Sonderegger}]{mcauliffe2017montreal}
Michael McAuliffe, Michaela Socolof, Sarah Mihuc, Michael Wagner, and Morgan Sonderegger. 2017.
\newblock Montreal forced aligner: Trainable text-speech alignment using kaldi.
\newblock In \emph{Interspeech}, volume 2017, pages 498--502.

\bibitem[{Morise et~al.(2017)}]{morise2017harvest}
Masanori Morise et~al. 2017.
\newblock Harvest: A high-performance fundamental frequency estimator from speech signals.
\newblock In \emph{INTERSPEECH}, pages 2321--2325.

\bibitem[{Ramesh et~al.(2021)Ramesh, Pavlov, Goh, Gray, Voss, Radford, Chen, and Sutskever}]{ramesh2021zero}
Aditya Ramesh, Mikhail Pavlov, Gabriel Goh, Scott Gray, Chelsea Voss, Alec Radford, Mark Chen, and Ilya Sutskever. 2021.
\newblock Zero-shot text-to-image generation.
\newblock In \emph{International Conference on Machine Learning}, pages 8821--8831. PMLR.

\bibitem[{Shen et~al.(2023)Shen, Ju, Tan, Liu, Leng, He, Qin, Zhao, and Bian}]{shen2023naturalspeech}
Kai Shen, Zeqian Ju, Xu~Tan, Yanqing Liu, Yichong Leng, Lei He, Tao Qin, Sheng Zhao, and Jiang Bian. 2023.
\newblock Naturalspeech 2: Latent diffusion models are natural and zero-shot speech and singing synthesizers.
\newblock \emph{arXiv preprint arXiv:2304.09116}.

\bibitem[{Shi et~al.(2020)Shi, Bu, Xu, Zhang, and Li}]{AISHELL-3_2020}
Yao Shi, Hui Bu, Xin Xu, Shaoji Zhang, and Ming Li. 2020.
\newblock Aishell-3: A multi-speaker mandarin tts corpus and the baselines.
\newblock \emph{arXiv preprint arXiv:2010.11567}.

\bibitem[{Von~Helmholtz(1912)}]{von1912sensations}
Hermann Von~Helmholtz. 1912.
\newblock \emph{On the Sensations of Tone as a Physiological Basis for the Theory of Music}.
\newblock Longmans, Green.

\bibitem[{Wang et~al.(2023)Wang, Chen, Wu, Zhang, Zhou, Liu, Chen, Liu, Wang, Li et~al.}]{wang2023neural}
Chengyi Wang, Sanyuan Chen, Yu~Wu, Ziqiang Zhang, Long Zhou, Shujie Liu, Zhuo Chen, Yanqing Liu, Huaming Wang, Jinyu Li, et~al. 2023.
\newblock Neural codec language models are zero-shot text to speech synthesizers.
\newblock \emph{arXiv preprint arXiv:2301.02111}.

\bibitem[{Wang et~al.(2022)Wang, Wang, Zhu, Wu, Li, Xue, Zhang, Xie, and Bi}]{wang2022opencpop}
Yu~Wang, Xinsheng Wang, Pengcheng Zhu, Jie Wu, Hanzhao Li, Heyang Xue, Yongmao Zhang, Lei Xie, and Mengxiao Bi. 2022.
\newblock Opencpop: A high-quality open source chinese popular song corpus for singing voice synthesis.
\newblock \emph{arXiv preprint arXiv:2201.07429}.

\bibitem[{Wu et~al.(2023)Wu, Chen, Zhang, Hui, Berg-Kirkpatrick, and Dubnov}]{wu2023large}
Yusong Wu, Ke~Chen, Tianyu Zhang, Yuchen Hui, Taylor Berg-Kirkpatrick, and Shlomo Dubnov. 2023.
\newblock Large-scale contrastive language-audio pretraining with feature fusion and keyword-to-caption augmentation.
\newblock In \emph{ICASSP 2023-2023 IEEE International Conference on Acoustics, Speech and Signal Processing (ICASSP)}, pages 1--5. IEEE.

\bibitem[{Yang et~al.(2023{\natexlab{a}})Yang, Liu, Huang, Lei, Weng, Meng, and Yu}]{yang2023instructtts}
Dongchao Yang, Songxiang Liu, Rongjie Huang, Guangzhi Lei, Chao Weng, Helen Meng, and Dong Yu. 2023{\natexlab{a}}.
\newblock Instructtts: Modelling expressive tts in discrete latent space with natural language style prompt.
\newblock \emph{arXiv preprint arXiv:2301.13662}.

\bibitem[{Yang et~al.(2023{\natexlab{b}})Yang, Tian, Tan, Huang, Liu, Chang, Shi, Zhao, Bian, Wu et~al.}]{yang2023uniaudio}
Dongchao Yang, Jinchuan Tian, Xu~Tan, Rongjie Huang, Songxiang Liu, Xuankai Chang, Jiatong Shi, Sheng Zhao, Jiang Bian, Xixin Wu, et~al. 2023{\natexlab{b}}.
\newblock Uniaudio: An audio foundation model toward universal audio generation.
\newblock \emph{arXiv preprint arXiv:2310.00704}.

\bibitem[{Yao et~al.(2023)Yao, Yang, Lei, Ning, Hu, Pan, Yin, Zhou, Lu, and Xie}]{yao2023promptvc}
Jixun Yao, Yuguang Yang, Yi~Lei, Ziqian Ning, Yanni Hu, Yu~Pan, Jingjing Yin, Hongbin Zhou, Heng Lu, and Lei Xie. 2023.
\newblock Promptvc: Flexible stylistic voice conversion in latent space driven by natural language prompts.
\newblock \emph{arXiv preprint arXiv:2309.09262}.

\bibitem[{Yu et~al.(2024)Yu, Simig, Flaherty, Aghajanyan, Zettlemoyer, and Lewis}]{yu2024megabyte}
Lili Yu, D{\'a}niel Simig, Colin Flaherty, Armen Aghajanyan, Luke Zettlemoyer, and Mike Lewis. 2024.
\newblock Megabyte: Predicting million-byte sequences with multiscale transformers.
\newblock \emph{Advances in Neural Information Processing Systems}, 36.

\bibitem[{Zeghidour et~al.(2021)Zeghidour, Luebs, Omran, Skoglund, and Tagliasacchi}]{zeghidour2021soundstream}
Neil Zeghidour, Alejandro Luebs, Ahmed Omran, Jan Skoglund, and Marco Tagliasacchi. 2021.
\newblock Soundstream: An end-to-end neural audio codec.
\newblock \emph{IEEE/ACM Transactions on Audio, Speech, and Language Processing}, 30:495--507.

\bibitem[{Zhang et~al.(2022{\natexlab{a}})Zhang, Li, Wang, Deng, Liu, Ren, He, Huang, Zhu, Chen et~al.}]{zhang2022m4singer}
Lichao Zhang, Ruiqi Li, Shoutong Wang, Liqun Deng, Jinglin Liu, Yi~Ren, Jinzheng He, Rongjie Huang, Jieming Zhu, Xiao Chen, et~al. 2022{\natexlab{a}}.
\newblock M4singer: A multi-style, multi-singer and musical score provided mandarin singing corpus.
\newblock \emph{Advances in Neural Information Processing Systems}, 35:6914--6926.

\bibitem[{Zhang et~al.(2022{\natexlab{b}})Zhang, Cong, Xue, Xie, Zhu, and Bi}]{zhang2022visinger}
Yongmao Zhang, Jian Cong, Heyang Xue, Lei Xie, Pengcheng Zhu, and Mengxiao Bi. 2022{\natexlab{b}}.
\newblock Visinger: Variational inference with adversarial learning for end-to-end singing voice synthesis.
\newblock In \emph{ICASSP 2022-2022 IEEE International Conference on Acoustics, Speech and Signal Processing (ICASSP)}, pages 7237--7241. IEEE.

\bibitem[{Zhang et~al.(2023{\natexlab{a}})Zhang, Liu, Lei, Chen, Yin, Xie, and Li}]{zhang2023promptspeaker}
Yongmao Zhang, Guanghou Liu, Yi~Lei, Yunlin Chen, Hao Yin, Lei Xie, and Zhifei Li. 2023{\natexlab{a}}.
\newblock Promptspeaker: Speaker generation based on text descriptions.
\newblock \emph{arXiv preprint arXiv:2310.05001}.

\bibitem[{Zhang et~al.(2022{\natexlab{c}})Zhang, Zheng, Li, and Lu}]{zhang2022wesinger}
Zewang Zhang, Yibin Zheng, Xinhui Li, and Li~Lu. 2022{\natexlab{c}}.
\newblock Wesinger: Data-augmented singing voice synthesis with auxiliary losses.
\newblock \emph{arXiv preprint arXiv:2203.10750}.

\bibitem[{Zhang et~al.(2023{\natexlab{b}})Zhang, Zheng, Li, and Lu}]{zhang2023wesinger}
Zewang Zhang, Yibin Zheng, Xinhui Li, and Li~Lu. 2023{\natexlab{b}}.
\newblock Wesinger 2: fully parallel singing voice synthesis via multi-singer conditional adversarial training.
\newblock In \emph{ICASSP 2023-2023 IEEE International Conference on Acoustics, Speech and Signal Processing (ICASSP)}, pages 1--5. IEEE.

\end{thebibliography}

\appendix
\section{Sample Prompt Keywords and Sentence Templates}
\label{appendix:prompt}

We list the keywords for each category in table \ref{tab:prompt_key}, and provide some samples of prompt sentence templates in table \ref{tab:prompt_sentence}.

\begin{table}[ht]
    \small
    \centering
    \begin{tabular}{l|l}
    \toprule
    Category & Keywords  \\ 
    \midrule
    \multicolumn{2}{l}{\textbf{Gender}} \\
    \midrule
    female & woman, lady, girl, female, lass, miss, madam \\
    \midrule
    male & man, boy, guy, gentleman, male, sir \\
    \midrule
    \multicolumn{2}{l}{\textbf{Volume}} \\
    \midrule
    \multirow{2}{*}{high} & loud, ringing, booming, thunderous, \\
    &  deafening, roaring \\
    \midrule
    \multirow{2}{*}{medium} &  moderate, average, intermediate, \\
    & middle-range \\
    \midrule
    low  & quiet, slight, twittering, hushed, whispering \\
    \midrule
    \multicolumn{2}{l}{\textbf{Vocal Range}} \\
    \midrule
    \multirow{2}{*}{high} & sharp, treble, shrill, whistling,\\
    &  shrieking, high-pitched \\
    \midrule
    low & deep, low, bass, thick, low-pitched \\

    \bottomrule 
    \end{tabular}
    \caption{Prompt keywords for each category.}
    \label{tab:prompt_key}
\end{table}

\section{Dataset Statistics}
\label{appendix:dataset}

In table \ref{tab:data_sta}, we list the statistics of the datasets used. F and M in the Speakers column indicate the numbers of female and male speakers or singers.

\begin{table}[ht]
    \small
    \centering
    \begin{tabular}{lcc}
    \toprule
    Dataset & \ Hours & \ Speakers \\ 
    \midrule
    \multicolumn{3}{l}{\textbf{SVS datasets}} \\
    \midrule
    M4Singer \cite{zhang2022m4singer} & 29.8 & F:10 M:10  \\
    Opencpop \cite{wang2022opencpop} & 5.3 & F:1  \\ 
    Opensinger \cite{huang2021multi} & 86.5 & F:49 M:28 \\
    PopCS \cite{liu2022diffsinger} & 5.9 & F:1 \\
    \midrule
    \multicolumn{3}{l}{\textbf{TTS datasets}} \\
    \midrule    
    AISHELL-3 \cite{AISHELL-3_2020}  & 86.4 & F:176 M:42 \\
    Biaobei \tablefootnote{https://www.data-baker.com/open\_source.html} & 11.8 & F:1   \\ 
    THCHS-30 \cite{THCHS30_2015} & 34.2 & F:31 M:9 \\
    Didispeech \cite{guo2021didispeech} & 47.0 & F:198 M:202 \\

    \bottomrule 
    \end{tabular}
    \caption{Statistics of training datasets.}
    \label{tab:data_sta}
\end{table}

\begin{table*}[t]
    \small
    \centering
    \begin{tabularx}{\textwidth}{X}
        \toprule
         \textbf{Single-Attribute Templates} \\
         \midrule
         Do you have any songs with a [gender] lead singer? \\
         Can you create a song sung by a [gender] vocalist? \\
         I'm searching for a song featuring a [gender] singer. \\
         I need a song with a [volume] voice that resonates.\\
         Play me a song with a [volume] voice. \\
         I'd like to listen to a song with a [volume] voice. \\
         I need a song where every note is gentle and delicate. (for low volume) \\
         Kindly provide me with a song that features a voice of balanced volume, pleasing to the ears. (for medium volume) \\
         Give me a song with a voice that shakes the ground with its thunderous vocals! (for high volume) \\
         \midrule
         \textbf{Double-Attribute Templates} \\
         \midrule
         Can you find me a song with a [gender] singer and a [volume] voice? \\
         I would like to hear a song with a [volume] voice and if possible, a [gender] voice. \\
         Synthesize a new song with a [volume] voice and a [gender] lead singer. \\
         Need a [pitch] pitch song sung by a [gender] vocalist. \\
         Generate a song featuring a [gender] vocalist with a unique use of [pitch] pitch. \\
         A [gender] voice with a [pitch] pitch is what I'm looking for. \\
         Create an enchanting song sung by a [gender] vocalist in the [pitch] pitch. \\
         Create a [gender] artist's song with a [volume] voice, softly mesmerizing with its gentle tone. (for low volume + any gender) \\
         Generate a [gender] artist singing at just the right volume. (for medium volume + any gender) \\
         Can you generate a [gender]-sung song with a [volume] voice that balances softness and loudness? (for medium volume + any gender) \\
         I'm looking for a song with a [gender] singer and a voice that's as powerful as a thunderstorm. (for high volume + any gender) \\
         \midrule
         \textbf{Triple-Attribute Templates} \\
         \midrule
         Explore [gender] [volume] songs with emotive [pitch] pitch. \\
         Synthesize a song with a [pitch] pitch and a [volume] voice, preferably [gender]. \\
         Design a [gender] singer's song with a [volume] voice and [pitch] pitch. \\
         Showcasing superb [pitch] pitch, create a [volume] song by a [gender] artist. \\
         Generate a song with stunning [pitch] harmonies and a [gender] singer with a [volume] voice. \\
         Can you compose a song with a [gender] vocalist and [volume] volume, while incorporating the singer's unique use of [pitch] pitch? \\
         Generate a song featuring [gender] vocals, delicately whispered with [volume] voice and [pitch] harmony. (for low volume + any gender / vocal range) \\
         Compose a [pitch]-keyed song with a [volume] voice that balances softness and loudness, sung by a [gender] singer. (for medium volume + any gender / vocal range) \\
         Craving a [gender] artist's song with a [volume] voice that exudes energy and power and a [pitch] note that creates a memorable hook! (for high volume + any gender / vocal range) \\
         \bottomrule
    \end{tabularx}
    \caption{Sample prompt sentence templates.}
    \label{tab:prompt_sentence}
\end{table*}
\section{Model Settings}
\label{appendix:model}

\begin{figure}[t]
    \centering
    \includegraphics[width=\columnwidth]{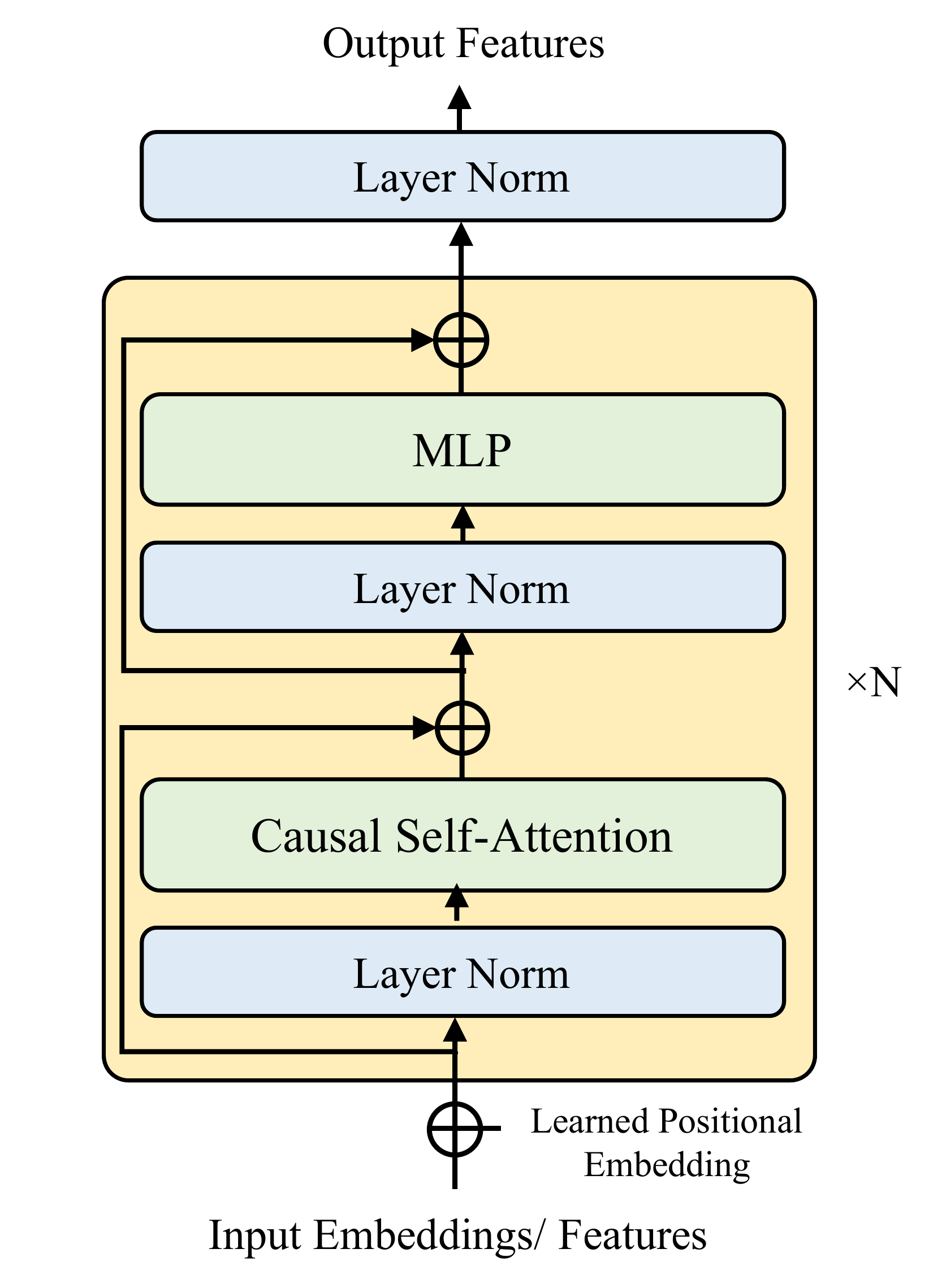}
    \caption{Structure of Global Transformer}
    \label{fig:global}
\end{figure}

We illustrate the architecture of the global transformer in Figure \ref{fig:global}. The local transformer shares the same structure as the global one with two differences: 1) the local transformer has no positional embedding, and 2) there is a linear lm-head appended to the top of it for token prediction. We also list the model hyper-parameters of Prompt-Singer in Table~\ref{tab:hyperparameters}. The multi-scale transformer is trained with 6 NVIDIA-V100 gpus for about 4-5 days, and the vocoder is trained with 4 NVIDIA-V100 gpus for a week.

\begin{table}[ht]
    \small
    \centering
    \begin{tabular}{c|c|c}
    \toprule
    \multicolumn{2}{c|}{Hyperparameter}   & Prompt-Singer \\ 
    \midrule
    \multirow{5}{*}{\tabincell{c}{Global\\Transformer}} 
    &Layers         &    20     \\
    &Hidden Dim                   &   1,152  \\    
    &Attention Headers           &  16   \\  
    &FFN Dim              &  4,608   \\    
    &Number of Parameters               & 320.07M \\
    \midrule
    \multirow{5}{*}{\tabincell{c}{Local\\Transformer}} 
    &Layers         &    6     \\
    &Hidden Dim                   &   1,152  \\    
    &Attention Headers           &  8   \\  
    &FFN Dim              &  4,608   \\    
    &Number of Parameters               & 100.13M \\
    \midrule
    \multirow{4}{*}{\tabincell{c}{Unit\\Vocoder}} 
    &Upsample Rates                     &    [6,5,2,2,2,2]     \\
    &Hop Size                 &   480  \\    
    &Upsample Kernel Sizes          &  [12,9,4,4,4,4]   \\
    &Number of Parameters               & 125.43M \\

    \bottomrule
    \end{tabular}
    \vspace{2mm}
    \caption{Hyperparameters of Prompt-Singer.}
    \label{tab:hyperparameters}
    \end{table}

\section{Evaluation Metrics}
\label{appendix:eval}
\subsection{Objective Evaluation}

For gender controlling accuracy, we train an open-source gender classifier\footnote{https://github.com/x4nth055/gender-recognition-by-voice/tree/master} with our singing and speech data. The performance of the classifier on the test set is provided as ground-truth accuracy in line 13 of table \ref{tab:res}.

For controlling accuracies on volume and vocal range, considering that the values of generated singing may slightly deviate from the boundaries used for categorization, we adopt a soft-margin mechanism for accuracy calculation. Specifically, 
we take the accuracy of data falling within the correct range as 100, and calculate the accuracy with $100 * \exp{(-k\epsilon)}$ for data outside the correct range, where $\epsilon$ is the error between the data value and the boundary, and $k$ is a hyper-parameter controlling the decay rate of accuracy at the margins, with larger $k$ corresponding to faster decay. We set $k$ to 10, 20 and 30 for high, medium and low volume, and 0.05 for vocal range accuracy.

\subsection{Subjective Evaluation}

For each evaluated model, we mix all generated results together and randomly select 220 items with their corresponding prompts for subjective evaluation. 

\begin{figure*}[tb]
\includegraphics[width=\textwidth]{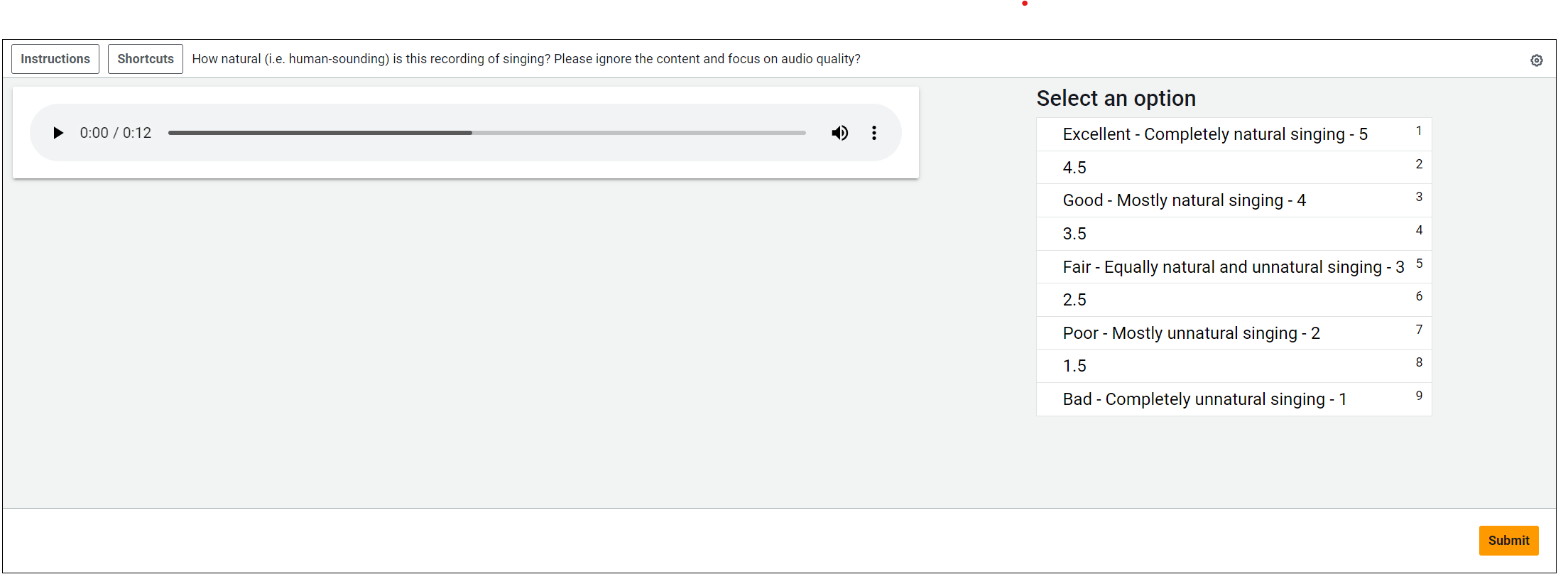}
\caption{Screenshot of MOS testing.}
\label{fig:mos}
\end{figure*}

\begin{figure*}[tb]
\includegraphics[width=\textwidth]{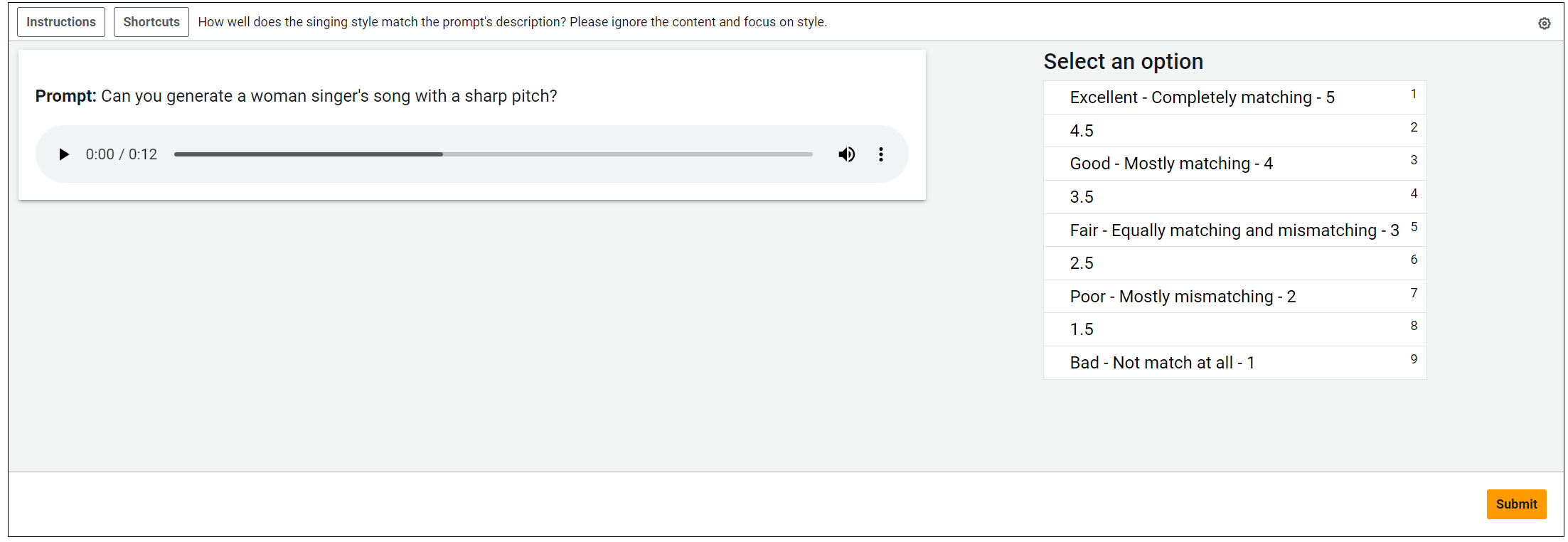}
\caption{Screenshot of RMOS testing.}
\label{fig:rmos}
\end{figure*}

Our subjective evaluation tests are crowd-sourced and conducted via Amazon Mechanical Turk. For audio quality evaluation, we ask the testers to examine the audio quality and naturalness and ignore the content. For prompt-style relevance, we instruct the testers to evaluate the relevance between the natural language prompt and the singing style while ignoring the content. The testers rate scores on 1-5 Likert scales. We provide screenshots of the testing interfaces in Figure \ref{fig:mos} and \ref{fig:rmos}. Each data item is rated by 4 testers, and the testers are paid \$8 hourly.

\section{Inference Efficiency}
\label{appendix:latency}

\begin{figure*}[tb]
\centering
    \small
    \subfigure[Latency of acoustic unit generation]
    {
      \label{fig:latency_acoustic}
      \includegraphics[height=0.38\textwidth]{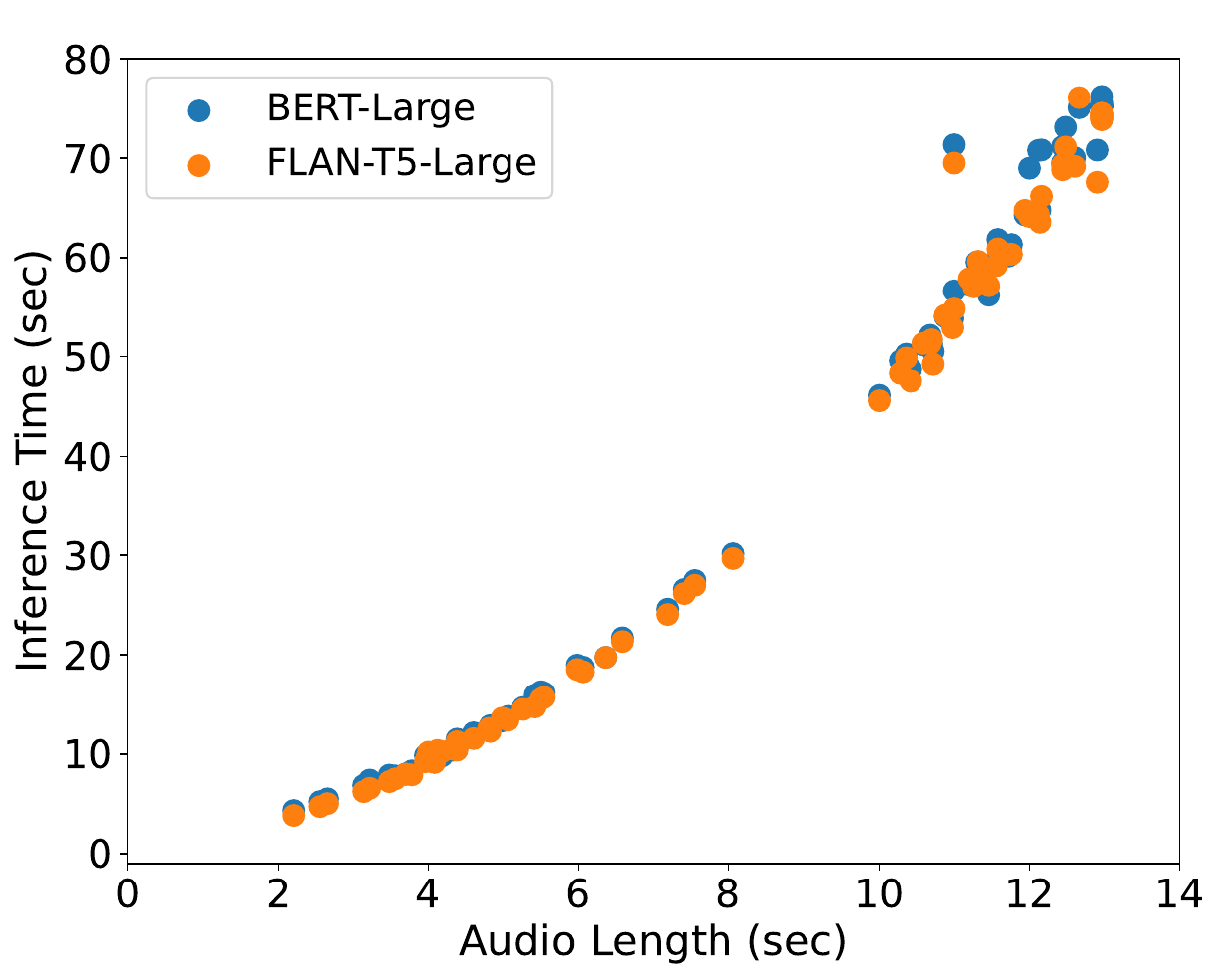}
    }
    \subfigure[Latency of wave reconstruction]
    {
      \label{fig:latency_voc}
      \includegraphics[height=0.38\textwidth]{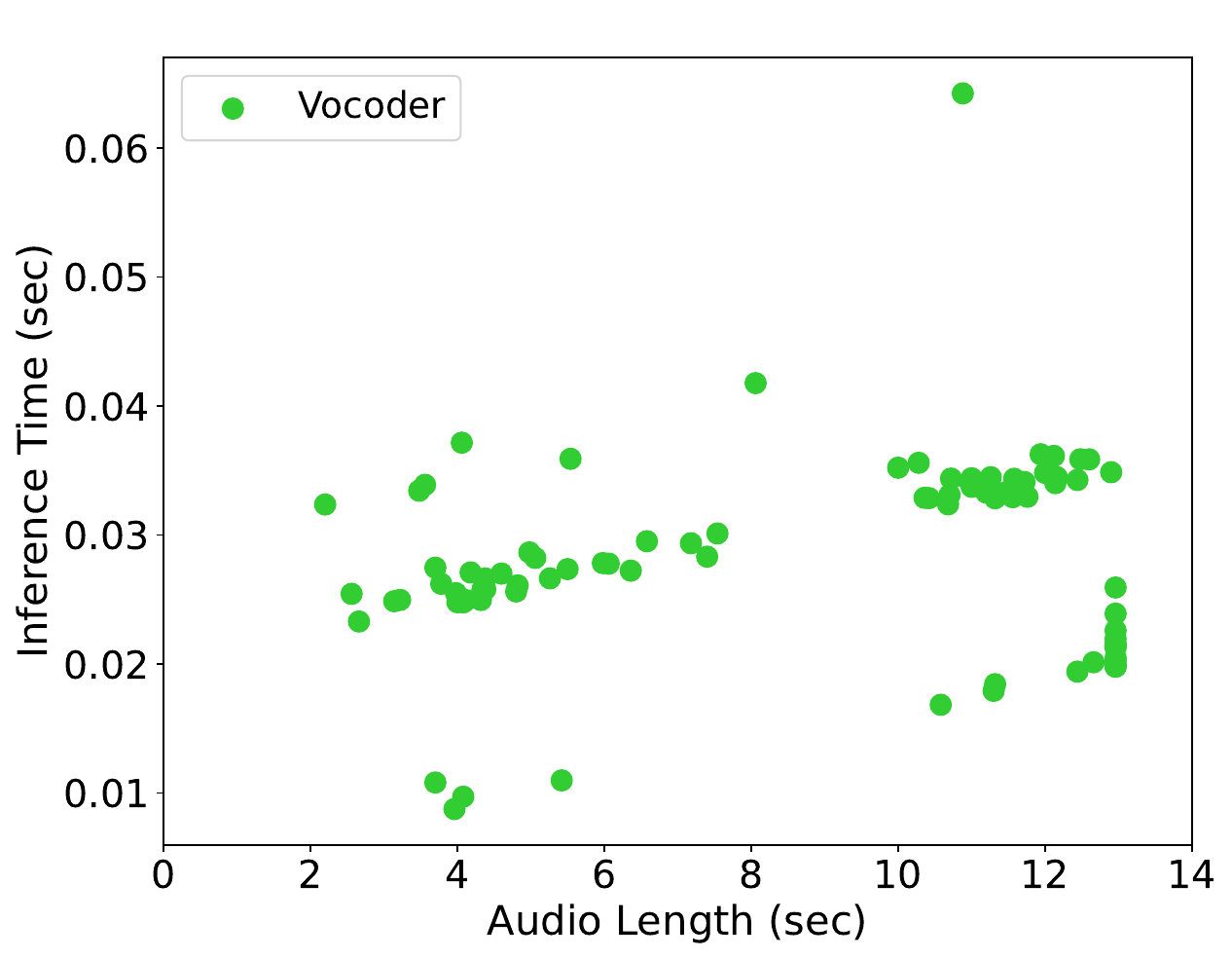}
    }
\caption{
Inference latency at varying lengths of generated audio.}
\label{fig:latency}
\end{figure*}

To give an intuitive impression of our model's inference efficiency, we visualize the relationship between model inference latency and the length of the generated audio in Figure~\ref{fig:latency}, including the acoustic unit generation stage with two types of text encoder, together with the wave reconstruction stage. The inference is conducted on a single NVIDIA-V100 GPU. It can be observed that the major latency comes from the transformer backbone, and it increases with the length of the sequence; on the other hand, the latency of the non-autoregressive vocoder is minimal and not significantly affected by the sequence length.

\end{document}